%% file: jem-euso-skeleton-bibtex.tex
\documentclass[a4paper,11pt]{article}
\usepackage{pos}

\usepackage{natbib}
\usepackage{subfig}
\title{Analysis of above-the-limb cosmic rays for EUSO-SPB2}
 \ShortTitle{Above-the-limb CR Analysis}

\author*[a]{Austin Cummings}

\affiliation[a]{Departments of Physics and Astronomy $\&$ Astrophysics, Institute for Gravitation and the Cosmos,\\
  Pennsylvania State University, University Park, PA 16802, USA}

\onbehalf{for the JEM-EUSO Collaboration\\[-1mm]{\normalsize \normalfont (a complete list of authors can be found at the end of the proceedings)}}
\emailAdd{alc6658@psu.edu}

\abstract{The Extreme Universe Space Observatory on a Super Pressure Balloon 2 (EUSO-SPB2)
experiment is a pathfinder mission for future space-based instruments targeting the fluxes of Ultra-High Energy Cosmic Rays (UHECR), with energies exceeding 1EeV and very high energy diffuse
and transient neutrinos, with energies exceeding 1PeV. Using two telescope designs: the
Fluorescence Telescope (FT) and the Cherenkov Telescope (CT), EUSO-SPB2 made novel
observations of the backgrounds relevant for space-based detection. EUSO-SPB2 will launch from
Wanaka, NZ in Spring of 2023, for a long duration (up to 100d) flight at a nominal float altitude of
33km.\\

In this contribution, we will focus on the CT’s capability to measure cosmic rays from above
Earth’s limb via the Cherenkov emission produced by the resultant Extensive Air Showers (EAS).
Using the \texttt{EASCherSim} optical Cherenkov generation code, we provide an updated estimate of the
event rate of above-the-limb cosmic rays for the CT, taking into account updated values for the
trigger efficiency as determined during the field testing of the instrument. We take particular care to
consider the longitudinal development of EAS in rarefied atmosphere, accounting for the energy
dependent elongation rate. In addition, we consider improvements to the magnetic field modeling
present in \texttt{EASCherSim} and illustrate their impact on the observed events and detection thresholds.
Finally, we compare these simulations to preliminary flight data from EUSO-SPB2.
}

\ConferenceLogo{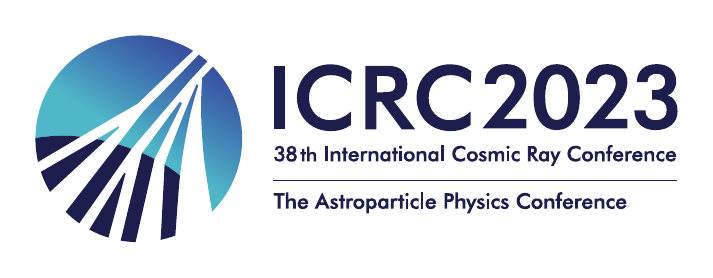}

\FullConference{%
38th International Cosmic Ray Conference (ICRC2023)\\
  26 July - 3 August, 2023\\
  Nagoya, Japan}

\begin{document}
\maketitle

\section{Introduction}
The Extreme Universe Space Observatory on a Super Pressure Balloon (EUSO-SPB2) is a pathfinder mission for the Probe of Multi-Messenger Astrophysics (POEMMA \cite{POEMMA-JCAP}), which aims to measure Ultra-High Energy Cosmic Rays (UHECR) and very high energy neutrinos from orbital altitudes. EUSO-SPB2 aims to test the measurement techniques of POEMMA by utilizing two telescopes in its design: i) the Fluorescence Telescope (FT), which points downward and measures fluorescence emission from UHECR induced Extensive Air Showers (EAS) and ii) the Cherenkov Telescope (CT), which points towards the Earth's limb and measures backgrounds for upward-going EAS sourced from $\nu_{\tau}$ interactions in the Earth. Although EUSO-SPB2 was not expected to observe the diffuse neutrino flux during its desired 100~d flight \cite{Cummings_2019}, significant work has been made to demonstrate its capability to perform transient follow-up measurements after an alert from a potential multimessenger event, considering the CT's unique capability to tilt and slew during flight \cite{Venters_2021, TobiasICRC}. In addition to measuring backgrounds and searching for neutrinos, the CT also tilts in zenith to measure above the Earth limb and observe emission from EAS sourced from atmosphere-skimming cosmic rays. Simulation work has estimated the rate of these above-the-limb cosmic rays to be hundreds of detected events per hour of live time, reaching energies as low as 1~PeV, helping to verify the detection technique and evaluate the instrument during flight \cite{Cummings_2021}. An overview of the EUSO-SPB2 instrument and mission can be found in \cite{HansICRC}.\\

\begin{figure}%
    \centering
    \subfloat {{\includegraphics[width=0.5 \linewidth]{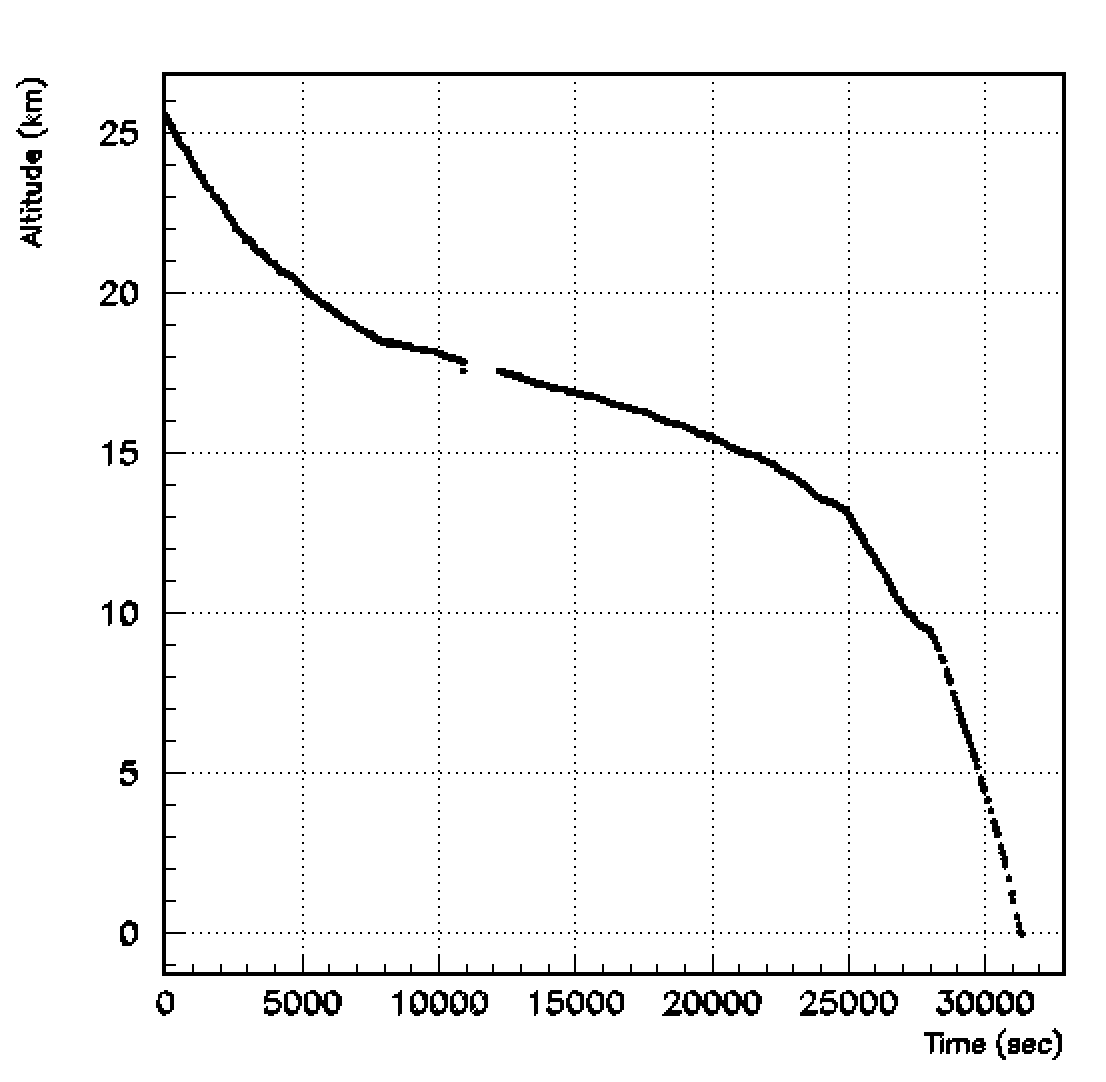} }}%
    \subfloat {{\includegraphics[width=0.5 \linewidth]{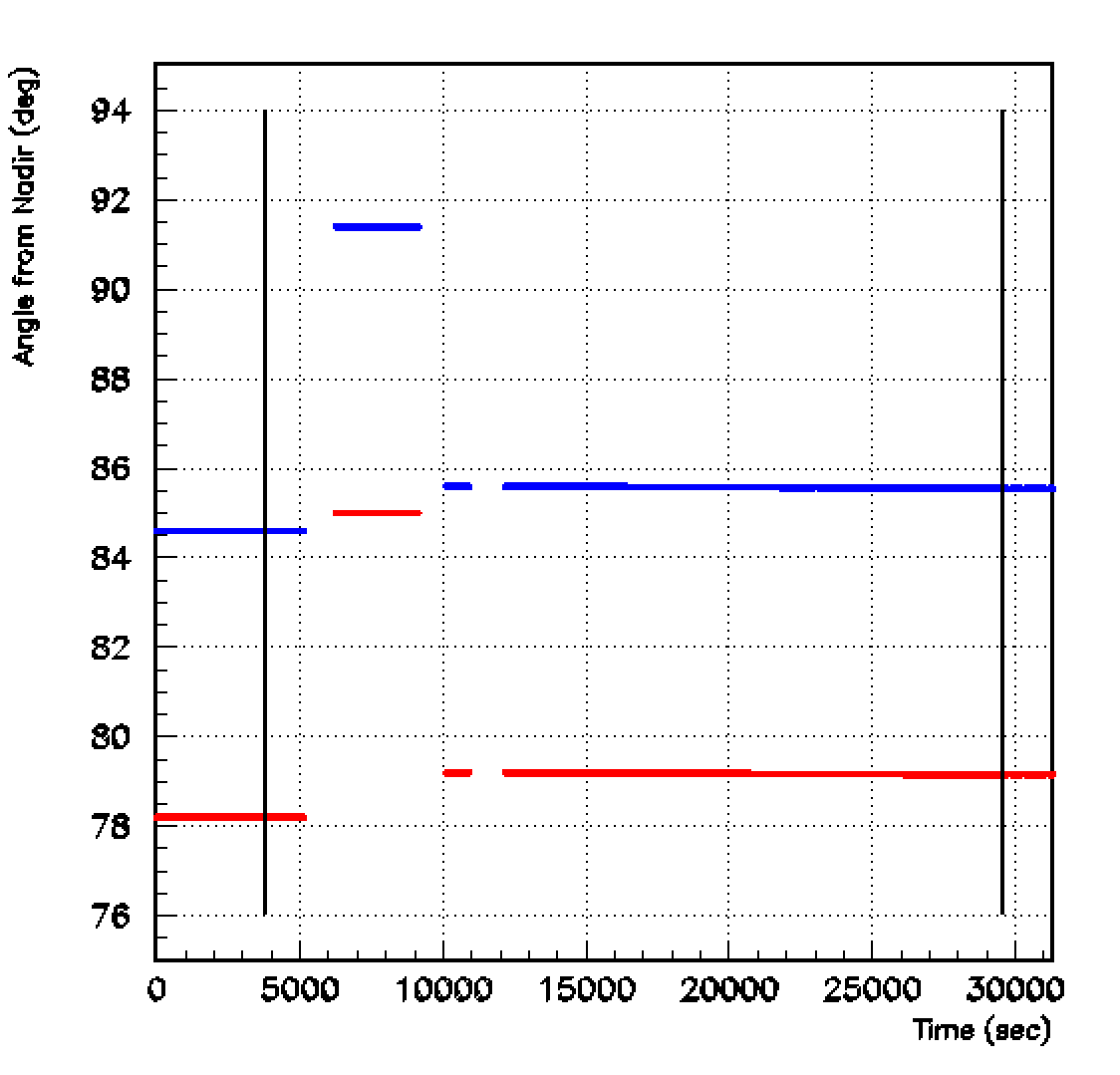} }}%
    \caption{Altitude (left) and viewing angle range (right) of EUSO-SPB2 during the duration of the flight. The blue and red lines in the right panel show the upper and lower edges of EUSO-SPB2's field-of-view. The shift at t = 6000~s corresponds to above-the-limb observation.}%
    \label{fig:flight}%
\end{figure}

EUSO-SPB2 launched from Wanaka, NZ on May 13, 2023 aboard a NASA Super Pressure Balloon and flew for 1 day, 12 hours, and 53 minutes before splashing down in the Pacific Ocean following an unexpected leak in the balloon. While this flight was too short for EUSO-SPB2 to meet its stated science goals, both instruments aboard the payload functioned as expected and managed to take data for much of the flight. The FT recorded and downloaded over 120,000 events over the two data-taking periods, which are currently under study to find signals consistent with an EAS \cite{GeorgeICRC}. While, for the majority of the flight, the CT made observations below the limb, providing novel background measurements for future space-based near-limb neutrino observation, roughly 45 minutes of data was taken above the limb. In total, 31,000 triggered events were recorded with the CT during flight, including several above-the-limb bi-focal triggers consistent with signals produced by above-the-limb cosmic rays following a preliminary analysis of the recorded data \cite{ElizaICRC}. The altitude and viewing angle range of EUSO-SPB2's CT over the course of the flight is shown in figure \ref{fig:flight}.\\

Previous estimates of the rate of above-the-limb cosmic rays were made assuming a float altitude of 33~km, and were lacking in their treatment of the geomagnetic field on produced Cherenkov emission. With the severely shortened time of flight for EUSO-SPB2, and the modified float altitude to an average of 18.5~km during the period of above limb observation, a new estimate is necessary to quantify the expected event rate of cosmic rays for the CT, folding in improved modeling of the geomagnetic field effects. This estimate will help guide the understanding of the recorded bi-focal events in the CT and serve to answer what their compatibility is with signals induced by cosmic ray EAS. In this work, we detail modifications made to the \texttt{EASCherSim}\footnote{https://c4341.gitlab.io/easchersim/} Cherenkov simulation code to consider the effects of geomagnetic spreading, recalculate the expected event rate of above-the-limb cosmic rays taking into consideration the flight characteristics, and interpret the observed event rate during flight in context of this calculation.

\section{Simulating Geomagnetic Effects}
\begin{figure}
\centering
\includegraphics[width=0.6\linewidth]{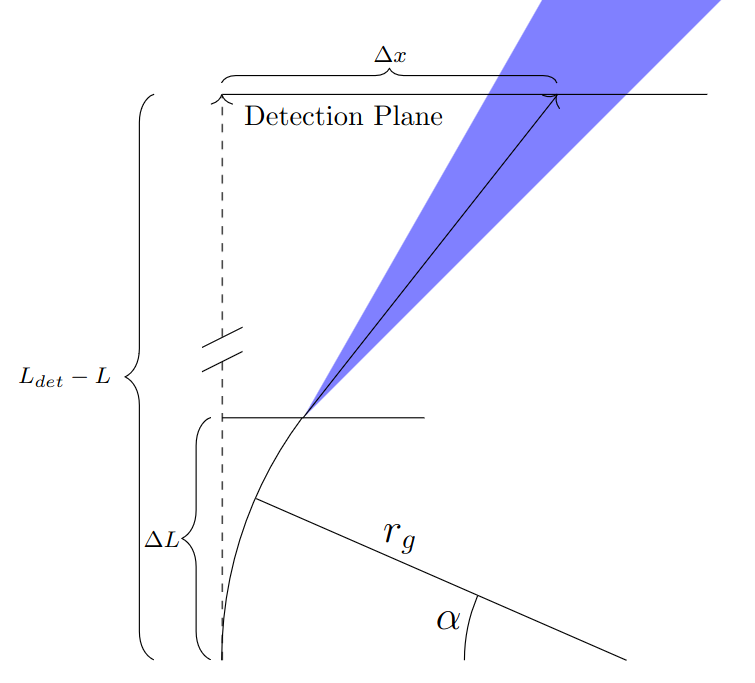}
\caption{Simplified diagram representing the geomagnetic deflection of electrons along the shower axis and the corresponding projection of the Cherenkov signal to an observing plane. See text for definitions of relevant variables.}
\label{fig:diagram}
\end{figure}

As extensive air showers evolve within Earth's atmosphere, charged particles are capable of being deflected by Earth's geomagnetic field via the Lorentz force. The electrons and positrons which comprise the bulk of the shower and produce the vast majority of the optical Cherenkov emission by the shower are light and therefore particularly susceptible to this deflection. We can model the effect of this deflection on the spatial distribution of arriving Cherenkov photons by making the assumption that an electron in the shower travels within a uniform magnetic field, with a circular arc of radius $r_{g}$, given by:

\begin{equation}
r_g = \frac{\gamma m_{e} \|\vec{v}\|}{q\| \hat{v} \times \vec{B} \|},
\label{eq:rg}
\end{equation}
where $\gamma$, $\vec{v}$, and $q$ are the Lorentz factor, velocity, and charge of the electron and $\vec{B}$ is the geomagnetic field. The magnetic field acts on the electron over a path length $\Delta L$ about the arc, and deflects the electron from a linear trajectory by the angle $\alpha = \Delta L/r_{g}$. Cherenkov photons are produced by the electron during the course of this deflection, and are thereby spread out on the detection plane. A simplified diagram of the effect of this deflection on the Cherenkov emission is shown in figure \ref{fig:diagram}.\\

To model the effective magnetic path length $\Delta L$ of an electron in the geomagnetic field, we use the parameterization given in \cite{Hillas:1982vn}:

\begin{align}
\langle \Delta L \rangle &= \frac{30 w^{0.2}}{1+36/E} \, (\mathrm{g}/ \mathrm{cm}^{2})\\
w &= 2(1-\mathrm{cos}\theta)(E/21)^{2}, \nonumber
\end{align}
where $E$ is the electron energy in MeV and $\theta$ is the electron angle with respect to the shower axis. At high $E$, $\langle \Delta L \rangle \sim 26 \, \mathrm{g}/\mathrm{cm}^2$, or about 0.7 electron radiation lengths. Concerning the spatial distribution of produced Cherenkov photons, magnetic deflection becomes a relevant process when the distance scale $\Delta L/\rho(z)$ and the gyroradius of an electron at the Cherenkov threshold $\big($which scales as $\rho(z)^{-1/2}$ $\big)$  are comparable, which occurs for altitudes around 25~km in air. For downward going EAS and upward going EAS induced by $\nu_{\tau}$ with energies below $10^{20}$~eV, maximum shower development occurs below these altitudes, and the effect of the geomagnetic field is minimal. However, for cosmic ray sourced EAS which develop above-the-limb, shower maximum is expected to occur above altitudes of 20~km, and this effect is nontrivial \cite{Cummings_2019, Cummings_2021}.\\ 

These effects are included in the \texttt{EASCherSim} computation scheme, taking into account electron properties and atmospheric effects, as well as the properties of the geomagnetic field. Above-the-limb EAS can often develop over hundreds of kilometers, experiencing a nontrivial variation in the geomagnetic field during flight. While modeling this variation is beyond the scope of this work, we can bound the maximal effects of the geomagnetic field on observed events by orienting the field perpendicular to the shower axis $\vec{L}$ and measuring the photon distributions along the axes parallel to $\vec{B}$ and $\vec{B} \times \vec{L}$. The angle of particles in the shower with respect to $\vec{L}$ are modified by $s \alpha$, where $s$ represents a value uniformly sampled between [-1, 1] to account for opposing deflections of electrons and positrons and for Cherenkov emission along the arc of travel (see figure \ref{fig:diagram}). While electron-positron charge asymmetry can maximally reach up to levels of 20\% in EAS, it is significantly smaller for electrons with energies characteristic for in-air optical Cherenkov emission. Thus, for what follows, we ignore the asymmetry.\\

\begin{figure}
  
\includegraphics[width=1\linewidth]{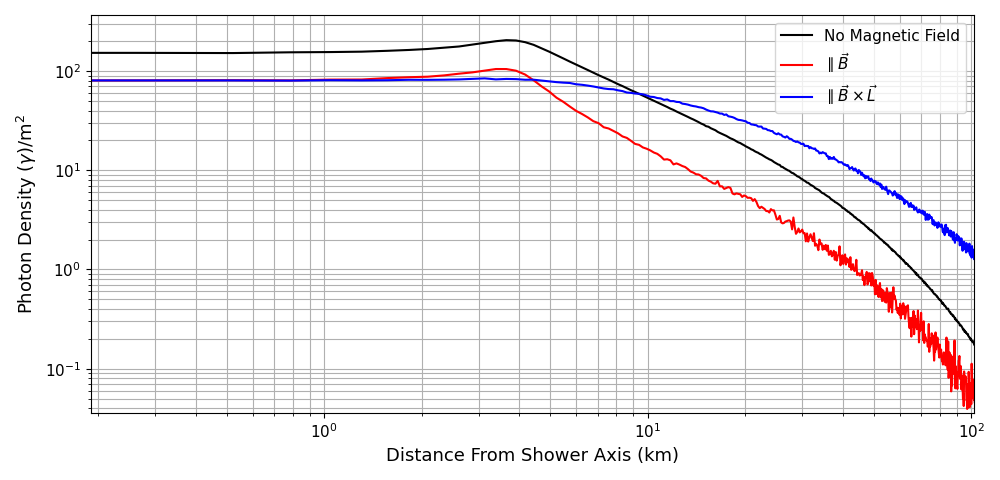}
\caption{Spatial distribution of Cherenkov photons at an altitude of 18~km produced by an above-the-limb EAS sourced from a 100~PeV proton with $\theta = 87^{\circ}$. The distribution is shown under the effect of no magnetic field and under the influence of a 50~$\mu$T field oriented perpendicular to shower development, measured along the axes parallel to $\vec{B}$ and $\vec{B} \times \vec{L}$.}

\label{fig:spatial_dist}
\end{figure}

Figure \ref{fig:spatial_dist} shows the effect of the geomagnetic field on the spatial distribution of Cherenkov photons for an example above-the-limb EAS for EUSO-SPB2. The Cherenkov photons are effectively spread along the axis perpendicular to $\vec{B}$, reducing the maximum photon density by roughly a factor of 2 and increasing the photon density off axis. This effect increases the energy threshold of detection and allows for greater probability of off-axis detection at high energies.

\section{Estimated Event Rate}

\begin{figure}%
    \centering
    \subfloat {{\includegraphics[width=0.5 \linewidth]{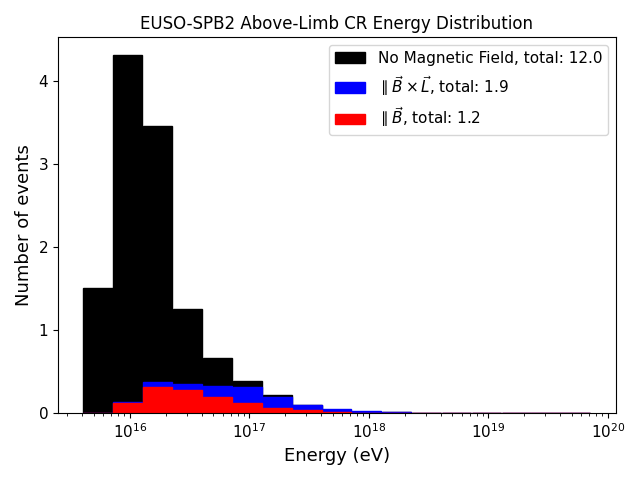} }}%
    \subfloat {{\includegraphics[width=0.5 \linewidth]{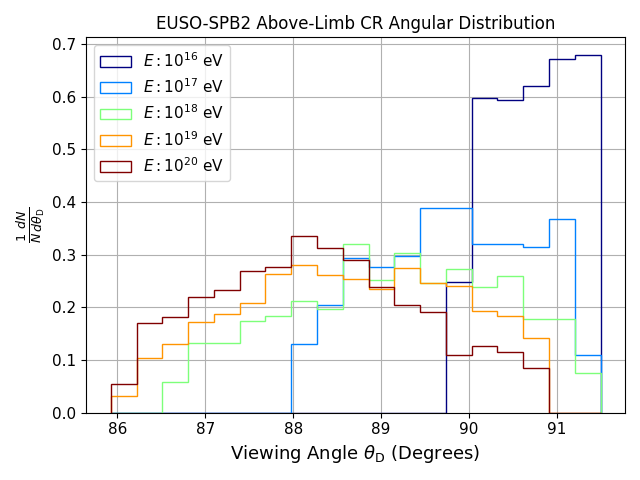} }}%
    \caption{(Left) EUSO-SPB2's estimated cosmic ray event rate of above-the-limb cosmic rays using the flight characteristics shown in figure \ref{fig:flight} and the flux measurements from the Pierre-Auger Observatory \cite{Aab:2019ogu}, calculated for quarter decade bins in energy. (Right) Normalized angular distribution of simulated above-the-limb cosmic rays as a function of energy, under the assumption of no magnetic field.}%
    \label{fig:area}%
\end{figure}

We simulate showers using \texttt{EASCherSim} for a detector altitude of 18.5~km and for viewing angles $85.7^{\circ}$ to $91.5^{\circ}$, representative of EUSO-SPB2 during the period of above limb observation. We take into consideration the energy dependence of the elongation rate by using the Gresien parameterization for $\gamma$-induced EAS and simulating showers in half decade energy bins from $10^{15}$~eV to $10^{20}$~eV. To bound the effects of the geomagnetic field on observed events, we simulate the properties of the Cherenkov emission under the configurations shown in figure \ref{fig:spatial_dist}. The trigger condition used in the full Monte Carlo simulation is taken from field test measurements performed in January 2022, and gives a threshold of of 150 photons for the $0.65~\mathrm{m}^2$ aperture over an integration time of 10~ns. The estimated event rate of above-the-limb cosmic rays as simulated for the 45 minute period of above-limb observation during flight is shown in figure \ref{fig:area}. The angular distribution of these simulated events is also shown as a function of energy.\\

Figure \ref{fig:area} shows that, compared to the estimates in \cite{Cummings_2021}, the cosmic ray energy threshold increases by roughly a factor of 3, resulting in a decrease in the observed number of events by a factor of $\sim 10$. The increase in threshold is primarily due to the decrease in altitude to 18.5~km from the desired 33~km. In the previous estimation, most of the observed events came from the upper end of the field of view (FOV), where the detector was positioned within active shower development, allowing for very bright signals. By descending to a lower altitude, the upper end of the FOV views portions of the atmosphere that are much thicker than those for the 33~km altitudes, resulting in dimmer signals. This can be observed in the angular distribution of simulated events via the cutoff in the distribution at high $\theta_{\mathrm{D}}$, particularly for low energy cosmic rays, where a significant amount of bright events occur outside the FOV.  Maximally applying the geomagnetic field to the simulated showers reduces the total event rate by roughly a factor of 6, due to the increased thresholds from the addition spreading. For cosmic ray energies $E>10^{18}$~eV, assuming showers are measured along the axes parallel to $\vec{B} \times \vec{L}$ and $\vec{B}$ yields event rates 2 times higher and 2 times lower than with no magnetic field present.\\

\begin{figure}
\centering
\includegraphics[width=1.\textwidth]{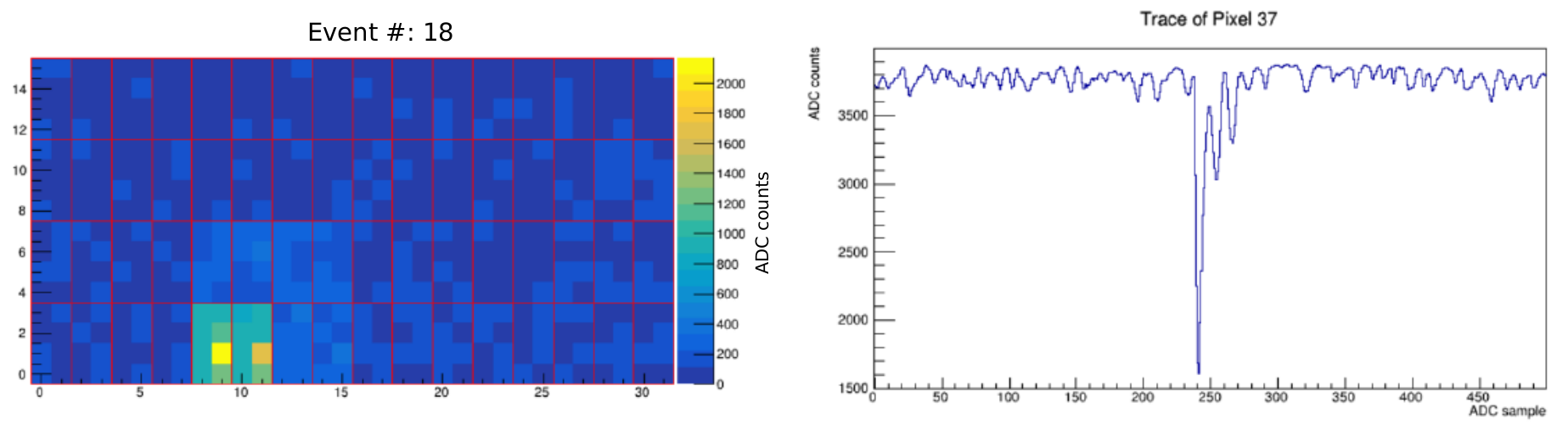}
\caption{Example above-the-limb cosmic ray candidate observed on May 14th 2023. (Left) single frame of the entire camera, where color represents the ADC counts. Note that lower pixels correspond to upper edge of FOV. (Right) time trace of the pixel with the highest signal.}
\label{fig:FlightCR}
\end{figure}

From preliminary analyses of the CT data, several bi-focal above-the-limb candidates consistent with cosmic ray signals have been found. Most of these events cluster near the upper edge of the FOV, and are quite bright with respect to background. An example above-the-limb cosmic ray candidate event recorded with the CT is shown in figure \ref{fig:FlightCR}. The observed event rate is consistent with the estimates presented here, being between $\sim 2$ and $\sim 12$ events, depending on the orientation of the geomagnetic field with respect to the incoming shower. The angular distribution of observed candidates also matches the simulated behavior shown in figure \ref{fig:area}.  

\section{Summary and Outlook}

EUSO-SPB2 was a mission of opportunity on a NASA super pressure balloon that aimed to evaluate the future POEMMA experiment's ability to detect UHECR and VHE neutrinos via the fluorescence and optical Cherenkov techniques. Although EUSO-SPB2 flew for a significantly shorter time than expected, all instruments aboard the payload operated as expected and recorded large amounts of data, helping to quantify backgrounds and instrument performance. For the CT, much of the observation time was spent recording near-limb backgrounds for future neutrino observations. For approximately 45 minutes at an altitude of $\sim$ 18.5~km, the CT was pointed above the limb to search for cosmic ray signals. Following preliminary analyses of the data, several bi-focal cosmic ray candidates were found \cite{ElizaICRC}.\\

A full Monte Carlo analysis of the above-the-limb cosmic rays using the \texttt{EASCherSim} Cherenkov simulation, taking into account the corrected effects of the geomagnetic field and the energy dependent elongation rate, estimates between 2 and 12 observed events during the above-limb data-taking period of the EUSO-SPB2 flight, depending on the shower orientation with respect to the geomagnetic field. It was noted that the shift between this new estimation and that for the 33~km flight (100 events/hour) primarily occurs due to the longer path lengths through the atmosphere provided by lower altitudes, resulting in dim signals. This estimation appears in line with the observations made by the CT, particularly considering that the observed backgrounds (and thereby thresholds) can be different than what was measured during the field tests. This finding is encouraging: should the observed events be cosmic rays (and they seem to consistent in both rate and signal characteristics), the measurement technique is validated. Further, we expect the rate of above-limb cosmic rays to increase significantly for future missions which fly at higher altitudes, providing a guaranteed signal by which to characterize the instrument.\\

Further analyses will be performed on the data successfully measured by both of EUSO-SPB2's instruments over the coming months, and will help to guide the observation strategy of the next pathfinder ballon mission: POEMMA Balloon with Radio. This experiment will be designed following the experience of EUSO-SPB2, but uniting the 2 optical telescopes into a single focal plane, similar to the original POEMMA design, with an added radio instrument to further improve detection rates and event quality. Following the analysis presented here, the telescope should be capable of recording hundreds of above-limb cosmic rays per hour of live time, and will help to optimize design parameters for future space-based missions. 
\input{Acknowledgement}
\bibliography{my-bib-database}

\input{JEM-EUSO_Authors_July2023.tex}



%
%
%

\end{document}

%% file: Acknowledgement.tex
\section*{Acknowledgment}
\noindent
\small{
The authors would like to acknowledge the support by NASA award 11-APRA-0058, 16-APROBES16-0023, 17-APRA17-0066, NNX17AJ82G, NNX13AH54G, 80NSSC18K0246, 80NSSC18K0473, 80NSSC19K0626, 80NSSC18K0464, 80NSSC22K1488, 80NSSC19K0627 and 80NSSC22K0426, by the French space agency CNES, and by National Science Centre in Poland grant n. 2017/27/B/ST9/02162. This research used resources of the National Energy Research Scientific Computing Center (NERSC), a U.S. Department of Energy Office of Science User Facility operated under Contract No. DE-AC02-05CH11231. We acknowledge the ASI-INFN agreement n. 2021-8-HH.0 and its amendments. We acknowledge the NASA Balloon Program Office and the Columbia Scientific Balloon Facility and staff for extensive support. We also acknowledge the invaluable contributions of the administrative and technical staffs at our home institutions.}

%% file: JEM-EUSO_Authors_July2023.tex
\newpage
{\Large\bf Full Authors list: The JEM-EUSO Collaboration\\}

\begin{sloppypar}
{\small \noindent
S.~Abe$^{ff}$, 
J.H.~Adams Jr.$^{ld}$, 
D.~Allard$^{cb}$,
P.~Alldredge$^{ld}$,
R.~Aloisio$^{ep}$,
L.~Anchordoqui$^{le}$,
A.~Anzalone$^{ed,eh}$, 
E.~Arnone$^{ek,el}$,
M.~Bagheri$^{lh}$,
B.~Baret$^{cb}$
% ...
}
\end{sloppypar}
\vspace*{.3cm}

{ \footnotesize
\noindent
% ...
$^{cb}$ Universit\'e de Paris, CNRS, AstroParticule et Cosmologie, F-75013 Paris, France\\
% ...
$^{ed}$ Istituto Nazionale di Fisica Nucleare - Sezione di Catania, Italy\\
% ...
$^{eh}$ INAF - Istituto di Astrofisica Spaziale e Fisica Cosmica di Palermo, Italy\\
% ...
$^{ek}$ Istituto Nazionale di Fisica Nucleare - Sezione di Torino, Italy\\
$^{el}$ Dipartimento di Fisica, Universit\`a di Torino, Italy\\
% ...
$^{ep}$ Gran Sasso Science Institute, L'Aquila, Italy\\
% ...
$^{ff}$ Nihon University Chiyoda, Tokyo, Japan\\ 
% ...
$^{ld}$ University of Alabama in Huntsville, Huntsville, AL, USA\\
$^{le}$ Lehman College, City University of New York (CUNY), NY, USA\\
% ...
$^{lh}$ Georgia Institute of Technology, USA\\
% ...
}

%% file: jem-euso-skeleton-bibtex.bbl
\providecommand{\href}[2]{#2}\begingroup\raggedright\begin{thebibliography}{10}

\bibitem{POEMMA-JCAP}
A.~Olinto et~al., \emph{The {POEMMA} ({Probe of Extreme Multi-Messenger
  Astrophysics}) observatory},
  \href{https://doi.org/10.1088/1475-7516/2021/06/007}{\emph{J. Cosmol.
  Astropart. Phys.} {\bfseries 2021} (2021) 007}.

\bibitem{Cummings_2019}
A.~Cummings, R.~Aloisio and J.~Krizmanic, \emph{Modeling of the tau and muon
  neutrino-induced optical {Cherenkov} signals from upward-moving extensive air
  showers}, \href{https://doi.org/10.1103/physrevd.103.043017}{\emph{Physical
  Review D} {\bfseries 103} (2021) }.

\bibitem{Venters_2021}
T.~Venters, M.H.~Reno and J.K.~and, \emph{Astrophysical implications of
  neutrino target-of-opportunity observations with space-based and suborbital
  optical {Cherenkov} detectors},  in \emph{Proceedings of 37th International
  Cosmic Ray Conference {\textemdash} {PoS}({ICRC}2021)}, Sissa Medialab, jul,
  2021, \href{https://doi.org/10.22323/1.395.0977}{DOI}.

\bibitem{TobiasICRC}
T.~Heibges et~al., \emph{{Overview of the EUSO-SPB2 Target of Opportunity
  program using the Cherenkov Telescope}}, {\emph{PoS} {\bfseries ICRC2023}
  (2023) }.

\bibitem{Cummings_2021}
A.~Cummings, R.~Aloisio, J.~Eser and J.~Krizmanic, \emph{Modeling the optical
  {Cherenkov} signals by cosmic ray extensive air showers directly observed
  from suborbital and orbital altitudes},
  \href{https://doi.org/10.1103/physrevd.104.063029}{\emph{Physical Review D}
  {\bfseries 104} (2021) }.

\bibitem{HansICRC}
J.~Eser et~al., \emph{{Overview and First Results of EUSO-SPB2}}, {\emph{PoS}
  {\bfseries ICRC2023} (2023) }.

\bibitem{GeorgeICRC}
G.~Filippatos et~al., \emph{{EUSO-SPB2 Fluorescence Telescope in flight
  performance and preliminary results}}, {\emph{PoS} {\bfseries ICRC2023}
  (2023) }.

\bibitem{ElizaICRC}
E.~Gazda et~al., \emph{{The EUSO-SPB2 Cherenkov Telescope - performance and
  preliminary results}}, {\emph{PoS} {\bfseries ICRC2023} (2023) }.

\bibitem{Hillas:1982vn}
A.~Hillas, \emph{{Angular and Energy Distributions of Charged Particles in
  Electron Photon Cascades in Air}},
  \href{https://doi.org/10.1088/0305-4616/8/10/016}{\emph{J. Phys. G}
  {\bfseries 8} (1982) 1461}.

\bibitem{Aab:2019ogu}
\emph{{The Auger Observatory: Contributions to 36th ICRC}}, {Sep}, 2019.

\end{thebibliography}\endgroup
